# A New Approach to Stateless Model Checking of LTL Properties

Elaheh Ghassabani and Mohammad Abdollahi Azgomi[*]

Trustworthy Computing Laboratory, School of Computer Engineering,
Iran University of Science and Technology, Tehran, Iran

E-mail: ghasabani@comp.iust.ac.ir, azgomi@iust.ac.ir

## Abstract

Verification of large and complicated concurrent programs is an important issue in the software world. Stateless model checking is an appropriate method for systematically and automatically testing of large programs, which has proved its power in verifying code of large programs. Another well-known method in this area is runtime verification. Both stateless model checking and runtime verification are similar in some ways. One common approach in runtime verification is to construct runtime monitors for properties expressed in linear temporal logic. Currently, there are some semantics to check linear temporal logic formulae on finite paths proposed in the field of runtime verification, which can also be applied to stateless model checking. However, existing stateless model checkers do not support LTL formulae. In some settings, it is more advantageous to make use of stateless model checking instead of runtime verification. This paper proposes a novel encoding of one of the recent LTL semantics on finite paths into an actor-based system. We take a truly parallel approach without saving any program states or traces, which not only addresses important problems in runtime verification, but can also be applied to stateless model checking.

## 1. Introduction

In recent years, it has become more prevalent to develop concurrent programs in order to utilize the computational power of parallel or multi-core processors. These types of programs are very difficult to verify. Even by using conventional methods of testing, such as various forms of stress and random testing, it is still difficult to detect all concurrency errors in such programs [1]. Obviously, it is not possible for programmers to manually verify concurrent programs with all their complexities; hence, automatic verification is an essential need in the software world. Software developers may use different methods to ensure their programs work correctly. A promising method for detecting and debugging concurrency errors [1, 2] is known as Model checking [2, 3].

---

[*] Corresponding author. Address: School of Computer Engineering, Iran University of Science and Technology, Hengam St., Resalat Sq., Tehran, Iran, Postal Code: 16846-13114, Fax: +98-21-73021480, E-mail: azgomi@iust.ac.ir.

In order for a programmer to directly verify code of a written program, *code model checking* [4, 5] is an appropriate method. From one point of view, code model checking can be classified into two categories: (1) stateful model checking, and (2) stateless model checking. Although stateful techniques are ideally suited to verify sequential programs, they usually run into the state space explosion problem verifying parallel programs. Owing to saving (all) the state space, the rise in the concurrency level may result in more complexity as well as the exponentially growth of the state space. In such situations, stateless model checking [1, 6] can be useful. Stateless model checking is especially appropriate to explore the state space of large and complicated programs because accurate capturing and controlling all the needed states of a large program could be a hard, or even impossible, task [1, 4, 7]. Global variables, heap, thread stacks, and register contexts are all part of the program state. Even if all the program states could be captured and controlled, processing such large states would be very expensive [8, 9].

A stateless model checker explores the state space of a program without capturing any program states. The program is executed under the control of a special scheduler, which systematically enumerates all execution paths of the program obtained by the nondeterministic choices. In other words, the scheduler controls the nondeterministic execution of threads [1, 4, 10]. As this method is applied to the source code level, it is very similar to software testing. In fact, it is a systematic testing method. A stateless model checker systematically explores all possible interleavings of the program threads under specific input for that program. That is to say, a stateless model checker explores the state space of a program by concretely and continuously re-executing the program such that the model checker generates a different thread scheduling scenario for each execution [11]. Therefore, all execution paths of the program generating by nondeterministic choices are covered [1, 10]. Although that stateless model checkers do not suffer from state space explosion owing to their stateless nature, they do not support verifying LTL formulae. However, there are some techniques for LTL checking in runtime verification [12] that can also be applied to stateless model checking.

Runtime verification is a technique in which, at run time, a monitor checks whether the execution of a system under inspection satisfies a given correctness temporal property [13]. Although runtime verification has a lot in common with stateless model checking, there is an important difference: in stateless model checking, all executions of a given system are examined to answer whether these satisfy a given property. In contrast, runtime verification does not consider each possible execution of a system, but just a single or a finite subset [14]. While both of the techniques are incomplete, stateless model it is not as incomplete as runtime verification. Even so, both of these techniques deal with finite traces and verify concrete executions of a program.

In runtime verification, the monitor interrupts the program execution when its state with respect to an event of interest changes, then evaluates a set of logical properties, and finally resumes the program execution. The main problem here is that the monitor has to act sequentially. Another problem is that, usually in runtime verification, the program trace first has to be stored and then can be analyzed [13].

Runtime verification of temporal logic properties requires a definition of the truth value of these properties on the finite paths that are observed at runtime. Although the semantics of temporal logic on infinite paths has been precisely defined, there is not yet an agreement on the definition of the semantics on finite paths [15]. Currently, there are some LTL semantics on finite paths [14-17], the most widely used of them for monitor construction are $LTL_3$ [14] and FLTL [16]. However, in [15], Morgenstern *et al.* proved that even a 4-valued semantics is not sufficient to achieve a semantics on finite paths that converges to the semantics on infinite paths. To the best of our knowledge, the semantic proposed in [15] is the most complete semantic that converges to the infinite path semantics [15].

This paper proposes a novel parallel encoding of LTL semantics [15] into an actor-based system, which can be used for monitor construction in runtime verification improving its efficiency. In addition, it is suited to apply to stateless model checking. In our method, there is no need to save any states or traces of a running program. Instead of translating an LTL formula into a Buechi automaton, which is the standard approach in model checking, the formula is translated into a set of actors that communicate with one another as well as with the main engine that explores the state space (i.e. the same stateless model checker). As state space explosion is one of the main obstacles in practical applications of model checking, having such techniques that do not rely on recording of the visited states, can be a solution to this problem. We model the proposed method using Rebeca [18, 19], which is an actor-based modeling language with a formal foundation.

The remainder of this paper is organized as follows. Section 2 gives the formal background required for this paper. Section 3 covers related work. Section 4 describes the proposed method for verifying LTL formulae. In this section, we model our method using Rebca modeling language, specify the properties of the model in LTL, and then describe the verification process and results. Section 5 gives an example to illustrate the proposed method. Section 6 briefly discusses the implementation issues. Finally, Section 7 mentions some concluding remarks.

## 2. Preliminaries

This section presents the formal background of this paper. The first subsection is a brief introduction to the semantics of LTL. The next subsection explains the semantic of a program state in stateless model checking. Finally, the last subsection briefly introduces the Actor model [20] as well as Rebeca modeling language [21] used in order to model actors' interactions.

### 2.1. Linear temporal logic (LTL)

This subsection is a brief introduction to (propositional) linear temporal logic [22], a logical formalism that is appropriate for specifying linear-time (LT) properties [2]. LTL is called linear because the qualitative notion of time is path-based and viewed to be linear: at each moment of time there is only one possible successor state, and thus each time moment has a unique possible future. Technically speaking, this follows from the fact that the LTL formulae are path-based (i.e. they are interpreted in terms of sequences of states) [2].

In the context of stateless model checking, we have to reason with linear temporal logic on truncated paths. A truncated path is a path that is finite, but not necessarily maximal [17]. Currently, a lot of different semantics for LTL on finite traces have been proposed [14, 16, 17, 23, 24]. As mentioned, we use the $RV^\infty$-LTL semantics [15] in this paper. So, this section describes the semantics of $RV^\infty$-LTL from [15].

For a given set of boolean variables (propositions) *AP*, the set of LTL formulas is defined by the following grammar: $\phi := a \mid \neg\phi \mid \phi \vee \phi \mid X \phi \mid [\phi \underline{U} \phi]$ where $a \in AP$. Additionally, $\phi \wedge \psi$, $F \phi$, $G \phi$, and *[ϕ U ψ]* are defined as abbreviations *for* $\neg(\neg\phi \vee \neg\psi)$, $[1 \underline{U} \phi]$, $\neg F \neg\phi$, and $[\phi \underline{U} \psi] \vee G \phi$, respectively. The semantics of LTL is usually given with respect to an infinite path through a transition system. These infinite paths are nothing else than infinite sequences of boolean assignments to the variables *a*:

**Definition 1.** (Infinite Words): Given a set of atomic propositions *AP*, an infinite word is a function $\sigma : N \rightarrow \wp(AP)$. For the sake of simplicity, $\sigma(i)$ is often denoted by $\sigma^{(i)}$ for $i \in N$. Using this notation, words are often given in the form $\sigma^{(0)} \sigma^{(1)} ....$ The suffix starting at *t* is written as: $\sigma^{(t...)} := \sigma^{(t)} \sigma^{(t+1)} ....$ For $a \in AP$, we define $\sigma = a^\omega$ as $\sigma = a^{(0)} a^{(1)} a^{(2)} ....$ Given an infinite word $\sigma = a^{(0)} a^{(1)} ...$, we define $\sigma^{(s...t)}$ as the finite word $u = \sigma^{(s)} \sigma^{(s+1)} ... \sigma^{(t)}$.

The semantics of LTL is typically defined as follows [15, 22]:

**Definition 2.** (Semantics of LTL on infinite paths). Given an infinite word $\sigma$, the following rules define the semantics of LTL:

- $[\sigma \vDash_\omega p]$ iff $p \in \sigma^{(0)}$ for $p \in AP$
- $[\sigma \vDash_\omega \neg \phi]$ iff $[\sigma \nvDash_\omega \phi]$
- $[\sigma \vDash_\omega \phi \wedge \psi]$ iff $[\sigma \vDash_\omega \phi]$ and $[\sigma \vDash_\omega \psi]$
- $[\sigma \vDash_\omega \phi \vee \psi]$ iff $[\sigma \vDash_\omega \phi]$ or $[\sigma \vDash_\omega \psi]$
- $[\sigma \vDash_\omega X \phi]$ iff $[\sigma^{(1...)} \vDash_\omega \phi]$
- $[\sigma \vDash_\omega [\phi \underline{U} \psi]]$ iff there is a $\delta$ such that $[\sigma^{(\delta...)} \vDash_\omega \psi]$ and for all $t$ with $t < \delta$, we have $[\sigma^{(t...)} \vDash_\omega \phi]$

In [15], a hierarchy of temporal formulae has been defined by the grammar rules of Table 1.

**Table 1.** Classes of the Temporal Logic Hierarchy [15]

| $P_G ::= p \mid \neg P_F \mid P_G \wedge P_G \mid P_G \vee P_G$ | $P_F ::= p \mid \neg P_G \mid P_F \wedge P_F \mid P_F \vee P_F$ |
|---|---|
| $\mid X P_G \mid [P_G U P_G]$ | $\mid X P_F \mid [P_F \underline{U} P_F]$ |
| $P_{Prefix} ::= P_G \mid P_F \mid \neg P_{Prefix} \mid P_{Prefix} \wedge P_{Prefix} \mid P_{Prefix} \vee P_{Prefix}$ ||
| $P_{GF} ::= P_{Prefix}$ | $P_{FG} ::= P_{Prefix}$ |
| $\mid \neg P_{FG} \mid P_{GF} \wedge P_{GF} \mid P_{GF} \vee P_{GF}$ | $\mid \neg P_{GF} \mid P_{FG} \wedge P_{FG} \mid P_{FG} \vee P_{FG}$ |
| $\mid X P_{GF} \mid [P_{GF} U P_{GF}] \mid [P_{GF} \underline{U} P_F]$ | $\mid X P_{FG} \mid [P_{FG} \underline{U} P_{FG}] \mid [P_G U P_{FG}]$ |
| $P_{Streett} ::= P_{GF} \mid P_{FG} \mid \neg P_{Streett} \mid P_{Streett} \wedge P_{Streett} \mid P_{Streett} \vee P_{Streett}$ ||

**Definition 3.** (Temporal Logic Classes): the logics $TL_\kappa$ for $\kappa \in \{G, F, Prefix, FG, GF, Streett\}$ is defined by the grammar rules given in Table 1, where $TL_\kappa$ is the set of formulas that can be derived from the non-terminal $P_\kappa$ ($p$ represents any variable $p \in AP$).

$TL_G$ is the set of formulae where each occurrence of a weak/strong until operator is positive/negative, and similarly, each occurrence of a weak/strong until operator in $TL_F$ is negative/positive. Hence, both logics are dual to each other, which means that one contains the negations of the other one. $TL_{Prefix}$ is the boolean closure of $TL_G$ and $TL_F$. The logics $TL_{GF}$ and $TL_{FG}$ are constructed in the same way as $TL_G$ and $TL_F$; however, there are two differences: (1) these logics allow occurrences of $TL_{Prefix}$ where otherwise variables would have been required in $TL_G$ and $TL_F$, and (2) there are additional 'asymmetric' grammar rules. It can be easily proved that $TL_{GF}$ and $TL_{FG}$ are also dual to each other, and their intersection strictly contains $TL_{Prefix}$.

Finally, $TL_{Streett}$ is the Boolean closure of $TL_{GF}$ and $TL_{FG}$. While there are syntactic restrictions on $TL_{Streett}$, i.e. not every LTL formula is a $TL_{Streett}$ formula, $TL_{Streett}$ contains for each LTL formula an equivalent formula, and nearly all formulas used in practice belong to $TL_{Streett}$. Moreover, for those formulas not in $TL_{Streett}$, it is typically not difficult to find an equivalent one in $TL_{Streett}$ [15].

### 2.1.1. Asymptotic Finite Linear Temporal Logic ($RV^\infty$–LTL)

Table 1 divides LTL formulae into different logics based on their grammar rules. The formulae whose specifications are based on the grammar $TL_\kappa$ fall into this type of logic. Practically speaking, we should always use the smallest logic, because of the quality of the results. How big a logic is, is determinable by Table 1: $TL_G$ and $TL_F$ are the smallest, then $TL_{Prefix}$, and so on. In order to verify a finite path, Morgenstern *et al.* [15] defined LTL semantics per each logic. Therefore, when it comes to the verification of a particular LTL formula, first it should be found out which category the formula belongs to. Then, the formula should be verified using the semantic of that category. [15]

**Definition 4.** (Semantics of Linear Temporal Logic $RV^\infty$–$TL_G$) [15]: Let $u = u^{(0)}u^{(1)}...u^{(n)} \in \Sigma^*$ denote a finite path of length $n + 1$. The truth value of a $TL_G$ formula $\phi$ wrt. $u$, denoted with $[u \vDash_G \phi]$, is an element of $B_3 = \{1, 0, \top_G\}$ and is inductively defined as follows:

- $[\varepsilon \vDash_G \phi] = \top_G$
- $[u \vDash_G a] = \begin{cases} 1 & \text{if } a \in u^{(0)} \\ 0 & \text{else} \end{cases}$ , for every $a \in AP$

- $[u \vDash_G \phi \wedge \psi] = \begin{cases} 1 & \text{if } \forall w \in \Sigma^\omega : uw \vDash_\omega \phi \wedge \psi \\ \top_G & \text{if } [u \vDash_G \phi] = \top_G \text{ and } [u \vDash_G \psi] = \top_G \\ 0 & \text{otherwise} \end{cases}$

- $[u \vDash_G \phi \vee \psi] = \begin{cases} 1 & \text{if } \forall w \in \Sigma^\omega : uw \vDash_\omega \phi \vee \psi \\ \top_G & \text{if } [u \vDash_G \phi] = \top_G \text{ or } [u \vDash_G \psi] = \top_G \\ 0 & \text{otherwise} \end{cases}$

- $[u \vDash_G X \phi] = [u^{(1...n)} \vDash_G \phi]$

- $[u \vDash_G [\phi U \psi]] = [u \vDash_G (\psi \vee (\phi \wedge X [\phi U \psi]))]$

**Definition 5.** (Semantics of Linear Temporal Logic $RV^\infty$–$TL_F$) [15]: Given a finite prefix $u = u^{(0)}u^{(1)}...u^{(n)}$ of an infinite word $u_\infty$, the semantics of $RV^\infty$–$TLF$ is defined by

- $[u \vDash_F \phi] = \begin{cases} 1 & \text{if } [u \vDash_G \neg\phi] = 0 \\ \bot_F & \text{if } [u \vDash_G \neg\phi] = \top_G \\ 0 & \text{otherwise} \end{cases}$

In [15], LTL semantics for all classes shown in Table 1 have been defined. In this paper, for the sake of brevity, we only propose our method for the classes $TL_G$ and $TL_F$. However, the method can be extended to cover all other semantics. It should be pointed out that the semantic of FLTL [16] and $RV^\infty$–$TL_G$ are evaluated in the same manner [15]. Therefore, our proposed method can be applied for monitor construction based on FLTL semantic as well. For a complete discussion on $RV^\infty$–$LTL$, please see [15].

## 2.2. Program states

In a multi-threaded program containing a finite set of threads and a set of shared objects, threads communicate with each other only through shared objects. Operations on shared objects are called visible operations, while the rest are invisible operations. A state of a multi-threaded program contains the global state of all shared objects and the local state of each thread. In a multi-threaded program, a visible operation performed by a thread is considered as a transition that advances the program from one global state to a subsequent global state. Such a transition is followed by a finite sequence of invisible operations of the same thread, ending just before the next visible operation of that thread [25, 26].

To avoid exploring redundant interleavings, stateless model checkers should use dynamic partial order reduction (DPOR) [26] because the number of possible interleavings grows exponentially as the program is getting large. Partial order reduction algorithms only explore a proper subset of the enabled transitions at a given state $s$ such that it is guaranteed to preserve the interested properties. DPOR dynamically tracks threads interactions to identify points where alternative paths in the state space need to be explored [11, 26].

To perform DPOR, a stateless model checker explores the program state space by concretely executing the program and observing its visible operations. It considers consecutive invisible operations with only one visible operation as a single operation [10, 27]. In this paper, we use the notion of code partitioning. Stateless model checkers are expected to apply some mechanisms for detecting global transitions. Therefore, we refer such mechanisms to partitioning, whereby code is divided into several global locations. In fact, the model checker interleaves threads according to these locations.

Each partition of the code (each location) starts with a visible operation, and ends just before the next visible operation. When a thread is scheduled, if it can progress, it continues executing until the end of its current location. After reaching the end of a location, it yields the processor to the model checker. If the thread holding the processor cannot progress, the scheduler should choose another thread. It goes without saying that this event may occur at the beginning of a location because only the first command of each location can be a waiting function call (a visible operation). Therefore, when it comes to LTL checking, we use the described definition for a state.

## 2.3. Actor model

*Actor* is a model for concurrent computing to develop parallel and distributed systems. Each actor is an autonomous entity that acts asynchronously and concurrently with other actors. It can send/receive messages to/from other actors, create new actors, and update its own local state. An actor system is composed of a collection of actors, some of whom may send messages to (or receive messages from) actors outside the system [28]. An actor using a command like *send(a, v)* creates a new message with receiver *a* and contents *v*, and then puts it to the message delivery system. This system guarantees the received message will be finally delivered to actor *a*. It can create another actor with a command like *newadr()*. Suchlike commands create a new actor and return its address. Each actor may have its own behaviors to process received messages. In other words, an actor's behavior embodies the code that should be executed by the actor after receiving a message [29].

As we stated, this paper uses the Actor model to propose its new verification method, which is also implemented by using an actor language. An actor language is an extension of a functional language. Erlang [30, 31] is arguably the best known implementation of the Actor model [28]. We are implementing the method proposed in this paper by using Erlang. In such languages, *functions* are used to define actors' behaviors. That is, each actor has a *behavioral functional* that embodies the actor behaviors after receiving particular messages.

In this paper, we model our method using Rebeca (Reactive Object Language) [18, 19], which is an actor-based modeling language with a formal foundation [18]. Then, we use the model checking technique to verify our models. For this purpose, model checker RMC [32] is used, which is a tool for direct model checking of Rebeca models, without using back-end model checkers. Using RMC, properties should be specified based on *state variables* of *rebecs*.

Rebeca is a Java-like language, which is mainly a modeling language with formal verification support and a background theory [21]. A Rebeca model consists of concurrently executing reactive objects called rebecs. In fact, rebecs are actors that communicate with each another by asynchronous message passing. Each message is put in the unbounded queue of the receiver rebec, specifying a unique method to be invoked when the message is serviced [19].

Figure 1 illustrates the definition of a simple Rebeca class. Although in a pure actor model the queue length is unbounded, the modeler has to declare the maximum queue size in the class definition owing to model checking. This size is indicated in parenthesis next to the *reactiveclass* name. A class definition uses two central declarations *knownrebecs* and *statevars*. The *knownrebecs* entry shows the actors this rebec can communicate with. The rebecs included in the *knownrebecs* part of a reactive class definition are those rebecs whose message servers may be called by instances of this reactive class. The *statevars* defines variables used for holding the rebec state [33].

After these declarations, the methods that handle messages are defined like Java code. These methods are called the message servers of the *reactiveclass* because their task is to serve incoming messages. Each reactive class definition has a message server named *initial*. In the initial state, each rebec has an initial message in its message queue, thus the first method executed by each rebec is the *initial* message server.

A message server contains one or more Rebeca statements. The logical and arithmetic expressions in Rebeca are similar to Java. However, not all of the Java expressions are valid in Rebeca, and only a set of essential set of operators are included [33]. For more information about Rebeca, please see [18, 33].

```
reactiveclass Rebec1(5) {
    knownrebecs { Rebec2 actor2 ; }
    statevars { }
    msgsrv initial ( ) {   self.msg1 ( );   }
    msgsrv serv_msg1( ) {
    /* Send a message to actor2, which should be processed by
      method process_msg in reactiveclass Rebec2. This message
      contains an integer value like "7"
    */
         actor2.process_msg (7);
    }
    msgsrv serv_msg2 ( ) {     /* Handling message 2 */      }
}
```

**Figure 1**    A typical class definition in Rebeca  [33]

The execution of rebecs in a Rebeca program takes place in a coarse grained interleaving scheme. In this manner, each rebec takes a message from the top of its queue and executes its corresponding message server. During execution, other rebecs are not allowed to be executed; i.e. the execution of a message server is atomic [33, 34].

## 3. Related work

As far as we know, prior to DSCMC [27, 35], there have been three stateless model checkers, namely Inspect [10, 36], CHESS [37, 38], and VeriSoft [4, 39]. Unfortunately, none of the foregoing tools support LTL formulae because, when it comes to the model checking field, LTL formulae reason about infinite traces. Stateless model checkers have no direct capability of reasoning about infinite traces. Stateless model checking is fairly a new trend in comparison with the stateful field. There is a lot of interesting work to do on this area. So, Extending stateless model checkers to check arbitrary LTL formulae is an interesting research direction.

LTL checking algorithms usually follow an automata-based approach taken from [40]. In this approach, the negation of the LTL formula is translated into a *Buchi automaton* [2, 41], synchronized with the transition system of the program state space, and then the verification problem is reduced to a simple graph problem [41]. Handling of large state spaces is so difficult (or even impractical) that the state space explosion has always been a pressing and serious problem in the stateful model checking field.

In order to verify LTL formulae, stateful model checkers have to capture the state space of the program, and the number of states grows exponentially in the number of variables in the program graph: for $N$ variables with a domain of $k$ possible values, the number of states grows up to $k^N$. Even if a program only contains a few variables, the state space that must be analyzed may be very large. This exponential growth in the number of parallel components and the number of variables leads to the enormous size of the state space of practically relevant systems. The reality is that verification problem in stateful model checking is particularly space-critical [2]. Nevertheless, many researches have been undertaken into this field leading the way to great achievements including some recent work in [42-45].

Of all the research in this area, the work by Ganai *et al.* [45] is more relevant to stateless model checking. Coping with state space explosion, they combined state-based and path-based (like stateless method) model checking, and then used a divide and conquer technique to explore state space. The main focus of their work is on proposing a new state exploration technique by combining state-based and path-based methods together. In other words, they also used the conventional techniques for verifying LTL formulae and did not propose a new LTL verification method (the focus of this paper).

Another work in this area was carried out by Evangelista and Kristensen [42]. They proposed an algorithm that is a combination of the common on-the-fly LTL model checking algorithms with sweep-line method [46]. Conventional on-the-fly LTL model checking is based on the exploration of a product Buchi automaton; i.e. the negation of the LTL formula to be checked is represented as a Buchi automaton, and then the product of this property automaton and the state space, viewed as a Buchi automaton, are explored using a nested depth-first traversal [41] in search for a cycle containing an acceptance state (an acceptance cycle). This work also has nothing to do with stateless model checking and is appropriate to the stateful techniques.

De Wulf *et al.* [43] proposed algorithms for LTL satisfiability and model-checking. In their algorithms nondeterministic automata were not constructed from LTL formulae. They directly alternated automata using efficient exploration techniques based on anti-chains. Similar to the previous work, their method is also not suitable for stateless model checking.

In the literature, the concept of runtime verification really stands out, which checks whether a system execution satisfies or violates a given correctness property [47]. A procedure that on-the-fly verifies conformance of the system's behavior to the specified property is called a monitor. The main idea of runtime verification is to monitor and analyze software and hardware system executions. Although this idea is fairly analogous to the idea of stateless model checking, methods used for runtime verification are completely different. In runtime verification, monitoring is carried out as follows. Two "black boxes", the system and its reference model, are executed in parallel and stimulated with the same input sequences; the monitor dynamically captures their output traces and tries to match them. The main problem is that a model is usually more abstract than the real system, both in terms of functionality and timing. For this reason, trace-to-trace matching is difficult, which causes the system to generate events in different order or even miss some of them [47].

Nowadays, there are a variety of formalisms to specify properties on observed behavior of computer systems including variants of temporal logic such as $LTL_3$ and TLTL [14]. In addition, currently, a lot of methods have been proposed to construct monitors [14, 15, 47]. As both stateless model checking and runtime verification deal with finite paths, current semantics of temporal logic on finite paths can be applied to the stateless model checking field.

The main problem in runtime verification is the extra overhead imposed by the monitor. Having an efficient monitor plays an important part in reducing the cost of verification. Several techniques have been proposed in this regard, such as improved instrumentation (e.g., using aspect-oriented programming [48]), combining static and dynamic analysis techniques [49, 50], efficient monitor generation and management[51], and schedulable monitoring [52]. Each of the forgoing approaches remedies the overhead issue to some extent and in specific contexts. However, there has been little work on reducing and containing the overhead of runtime verification through isolating the monitor in a different processing unit. With this motivation in mind, Berkovich *et al.* [13] proposed a technique that permits the separation of the functional from monitoring concerns into different computing units. Their formal language for monitoring properties is $LTL_3$. In their method, a host process receives the program trace in the shared memory and distributes chunks of this trace among a set of monitoring worker threads running on the GPU. The worker threads are capable of monitoring one or more properties simultaneously.

Our method can also isolate the monitoring unit in a different computing unit. In [13], the program trace has to be saved first, and then it is distributed to different processing units for analysis. However, in our method, there is no need to save any program traces or states at all. The problem with saving the program trace is that the causal order of occurrence of events in the program trace has to be respected while evaluating a property in a parallel fashion. For this reason, authors in [13] had to formalize a notion of $LTL_3$ property history to encode the causal order of events for parallelization. Therefore, their approach demands more time and memory in comparison with our proposed method.

To sum up, as far as we know, any LTL verification method in the stateless model checking field has not been proposed yet; this paper presents a new LTL checking method for this field, which can also be considered as a method for constructing a parallel and distributed monitor in runtime verification.

## 4. An Actor formalism for LTL

This section describes a novel method for stateless model checking of LTL formulae. In this method, LTL formulae are dynamically checked during program execution without storing any program states. For this reason, it is possible to verify any number of LTL properties away with affecting on the size of the program state space and state space explosion. The method to verify LTL formulae, proposed in this section, is quite different from conventional LTL checking algorithms.

In our method, we suppose that there is a stateless model checker that runs the program under test and systematically explores its state space. This model checker should accept all possible interleavings under strong fairness [1, 2, 27]. To generate different possible interleavings, the program must be repeatedly run under the stateless model checker until all possible thread scheduling options are generated. This paper concentrates on how LTL properties can be verified using such model checkers. Our method is proposed as a *unit of LTL checking*, which should cooperate with the stateless model checker. This unit is an actor system, a collection of actors with a hierarchical structure.

It should be pointed out that the process of stateless model checking is composed of finite *iterations*. Each iteration is equivalent to the execution of program $P$ under the control of the model checker scheduler. Program $P$ satisfies LTL property $\phi$ if $\phi$ is held by all iterations of stateless model checking. Intuitively, if an LTL property is violated in one program execution, it means that the property has been violated in one path of the program state space; consequently, the program does not satisfy this property. In the same way, if an LTL property is held by all iterations of stateless model checking then the property is satisfied by all paths in its state space; consequently, the program satisfies the property.

We use this idea as a basis for LTL checking in stateless model checking. It shows the feasibility of applying LTL checking to this field. But, the major need in this regard is to have an LTL checking method that can work with the stateless nature of the model checker. The remainder of this section proposes such method to solve this problem.

In light of the grammar of LTL formulae, terminals in this grammar are atomic propositions (i.e. $a \in AP$) [2]. An atomic proposition is a simple condition defined on program variables (e.g. *a > 0*, *b = 0*, *c != d*, etc.). Therefore, every LTL formula ends in simple conditions. The result of a simple condition is always either *true* or *false*. We use this fact for designing the unit of LTL checking.

Now, let us introduce the idea of the method with a simple example. Suppose you specify an LTL property as *"(¬ ((a > 0) ∧ (b = 1))) U (c = 0))"* where *a, b,* and *c* are integer variables in the program. The parse tree for this property is shown in Figure 2. All LTL properties, like this property, are evaluated from leaves towards the root of the parse tree; i.e. in this example, first, operator *and* (∧) should be evaluated, next, the *not* operator (¬), and then operator *until* (*U*) can be evaluated. We exploit this fact in our method; as it can be seen, leaves of a parse tree are simple conditions (or *APs*) while both of its root and intermediate nodes are LTL operators.

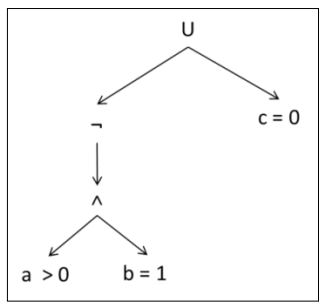

**Figure 2**  The parse tree of (¬ ((a > 0) ∧ (b = 1))) U (c = 0))

To check an LTL formula by our method, first a property is parsed, next an actor whose behavior corresponds to the root operator is created, and then existing sub-trees of the root are sent to the behavioral function of this actor as its arguments; e.g. in the above example, an actor who behaves corresponding to operator *U* is created and two sub-trees are sent to its behavioral function as its input arguments. Thereafter, this actor also makes the parse tree for each input argument (i.e. each sub-tree). In the same way, an actor for the root operator of each sub-tree is created and related sub-trees are sent to them. This process is continued by new actors until reaching the leaves; e.g., in this example, the actor with *until* behavior creates another actor with *not* behavior, and then the created actor creates a new actor with *and* behavior. When this new actor reaches a leaf (i.e. a simple condition) after parsing one of its arguments, it should create a new actor that checks a simple condition (we call such actors *condition checkers*). The intermediate nodes of the primary parse tree are called *workers* that are actors that behave corresponding to LTL operators. This hierarchical structure described here is shown in Figure 3 (a). There are two other kinds of actor in this hierarchy, *property checker* and *master*, which are described below.

In this paper, we suppose the existence of a mechanism in the model checker so that *condition checkers* are be able to monitor the state of the intended *APs*. At the implementation level, the model checker can think of different mechanisms. For example, based on the property the user has defined, it can instrument the program code such that at every point in the code that the variables in the *APs* of the property are defined[†], a piece of code is added to the original code, by which the simple conditions in the property (i.e. *APs*) can be monitored during stateless model checking. By doing so, *condition checkers* are informed about the status of their desired *APs* at the end of each state.[‡] A similar mechanism has been implemented in DSCMC [27].

---

[†] The variable definition means that a new value is assigned to the variable (*e.g.* using of the assignment operator *"="*).
[‡] The definition of a state in the context of stateless model checking is given in Section 2.1.3.

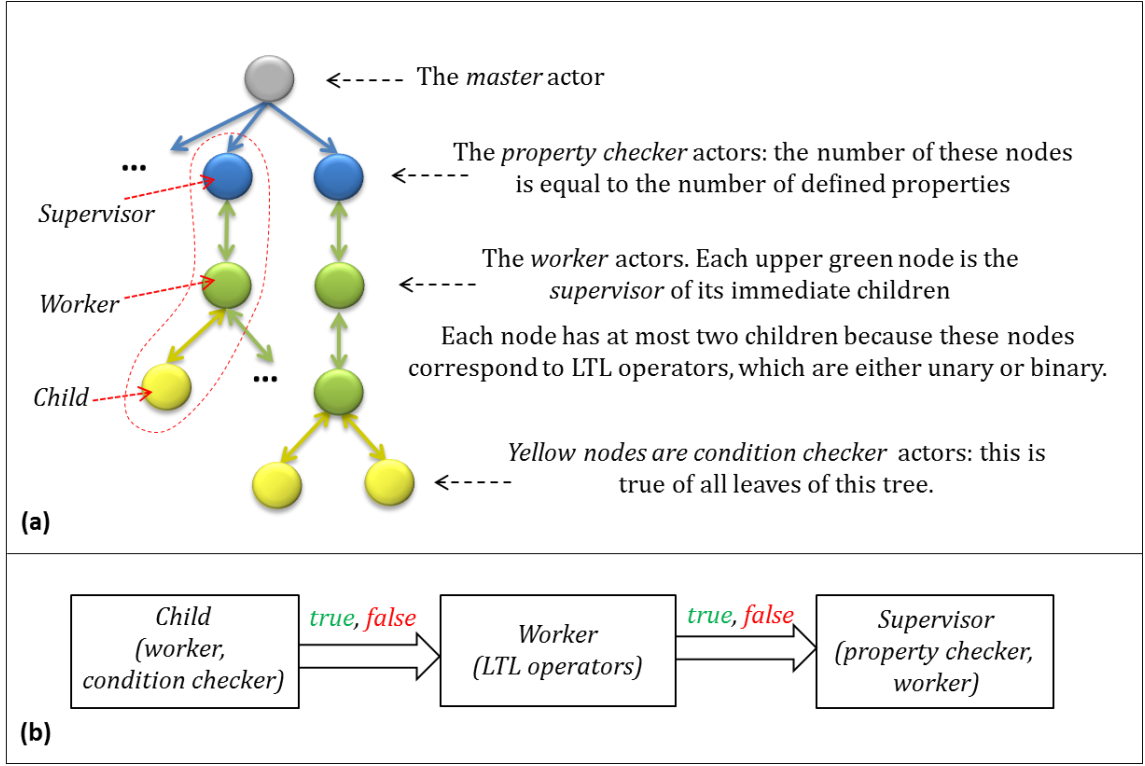

**Figure 3**  The hierarchical structure of the *Unit of LTL checking*. (a) Existing actors and their roles. (b) Usage of the hierarchical structure of the actor system for modeling.

The *unit of LTL checking* (Figure 3 (a)) has a major actor as the *master* actor, whose task is to load the user-defined LTL properties at the beginning of stateless model checking, and then create a *property checker* actor for each property. The *property checker* actors use a function for parsing a given property. This function creates the parse tree of its input argument, and then returns the root of this tree and sub-trees of the root. Thereafter, the *property checker* creates a *worker* that will be in charge of the sub-tree. The return sub-tree is also sent as an input argument to the behavioral functions of this *worker*.

As mentioned, the created *worker* by *property checker* also parses its input arguments (i.e. sub-tree(s) sent by *property checker*). Then, with respect to the parse tree of its arguments, it also creates other *worker* actor(s). Needless to say, *workers* are different in their behavior. Permissible behaviors for *workers* exactly correspond to LTL operators. For example, *and worker*, *not worker*, and *until worker* have the same semantics of operators $\wedge$, $\neg$, and $U$, respectively. You can see the procedure for creating the described hierarchy in Figure 4.

| **1.** master (Properties) {  | **2.** property checker (Property) { |
|---|---|
|    Load Properties; |    Parse Property; |
|    For each Property in the Properties list |    For the root operator |
|       Create a *property checker* actor; |       Create a *worker* actor with behavior of the root; |
|       Assign each *property checker* a Property; |       Assign the subtree of the root to the *worker*; |
| } | } |
| **3.** worker (Args) { | |
|    Parse each available subtree in Args; |    If there is no LTL operator in the root |
|    For each available root operator in subtrees of Args |       Create a *condition checker*; |
|       Create a *worker* actor with behavior of the root; |       Assign the related AP to the *condition checker*; |
|       Assign the subtree of the root to the *worker*; | } |

**Figure 4** Procedure for creating the verification hierarchy

In our method, each *property checker* actor parses an LTL property, and then creates a *worker* actor to evaluate the operator in the root of the parse tree. Consequently, the created *worker* also repeats this process until a *worker* actor reaches the simple condition(s). In other words, tasks are downwardly dispatched, then, results are upwardly collected from *workers* to their supervisors, and finally the results of evaluating get to *property checkers*, which are at the top level of the verification hierarchy (of course, after the *master*).

## 4.1. Modeling the stateless model checking of LTL operators in Rebeca

This section models the actor system described in the previous subsection. In this regard, we use Rebeca modeling language [18, 19].

We need an abstract model that correctly embodies the possible interactions between actors. For this purpose, we exploit the hierarchical structure shown in Figure 3. In this structure, the position of an intermediate node (*worker*) is similar to Figure 3 (b). As described earlier, *workers* correspond to LTL operators. The behavior of each *worker* actor is modeled in Rebeca using the structure shown in Figure 3. That is, each *worker* has a supervisor and at least one child. In other words, each *worker* is an LTL operator that can have at most two children. Each child may also be an LTL operator. Besides, *condition checkers* are also children of their immediate parent (*worker*). Each *worker* has one supervisor (its immediate parent), which may be an LTL operator or a *property checker*. As it can be seen, a child only sends its evaluation result (*true* or *false*) to its *supervisor*, regardless of whether it is a *condition checker* or an LTL operator. A *supervisor* also can receive either *true* or *false* from the *worker* regardless of the fact that the *worker* is which LTL operator.

On account of the above structure, it does not matter to a *worker* who its children and its supervisor are. Every *worker* only receives *true* or *false* from its children, and only sends *true* or *false* to its supervisor. Therefore, we can model the LTL checking unit, and verify the behavior of each actor (i.e. behavioral functions) independently. In this model, the behavior of each *worker* is characterized in Rebeca, and then other actors the *worker* can communicate with are modeled as black boxes that correspond to the same *worker's* supervisor and children. That is, black boxes used in the model, namely *Child*, and *Parent*, are actors that behave like a *child* and a *supervisor*, respectively (see Figure 3). A child is expected to send only either *true* or *false* to the *worker*, and a supervisor also expects to receive the result of the verification from the *worker*. Even so, in practice, when there are some until operators in a property, some actors may be waiting for receiving results from their children. In other words, it is possible for an intermediate actor not to hear from its children in a particular state. We have to model this fact, hence, in our models, *workers/ supervisors/ children* send three types of message: (1) a message that models *true*, (2) a message that models *false*, and (3) a message that models a waiting state.

Before moving on to modeling, we should point out the role of actors *master*, *property checker*, and *condition checker*. These actors are important when it comes to the implementation of the model. In terms of modeling, it makes no difference to the result of verification who creates *property checkers* and *workers*, or how the model checker informs *condition checkers* about the *APs* status. The main focus of the model should be maintained on how a *worker* behaves as a particular LTL operator, and how it evaluates its operands. For this reason, we model a *worker* regardless of who its parent and child (children) are. For instance, as for the *Until* operator, the model should demonstrate the way by which this actor evaluates results of its operand; e.g. how to act when it receives a *true* message from its left operand, how to act when it never receives a *true* from its right operand, and so on.

We model the behavior of the *workers* that correspond to *and (∧)*, *not (¬)*, and *Until (U)* operators. Other LTL operators can be derived from these operators. Each of the forgoing operators is independently modeled; i.e. the children (or supervisor) of each operator are viewed as a black boxes that only send (or receive) *true* and *false*.

Figure 5 models three rebecs that correspond to (a) stateless model checker, (b) *child*, and (c)

*supervisor*. The rebec who models stateless model checker, *SMC*, initiates the execution of the model from the method *sendAP* on line 6 of Figure 5 (a). This rebec models the fact that at the end of each state, the stateless model checker sends the status of the desired *AP*s to the *condition checkers*. After that, *condition checkers* send the results to their supervisors, and next their supervisors, according to their own functionality, evaluate the results and send them to their own supervisors. In practice, this process should continue until the most upper *worker* evaluates the results and sends the result of its evaluation to its own *property checker*. The property checker has to come a decision about the verification result at the end of each iteration. As you can see, the end of each iteration has been modeled at line 9 of Figure 5 (a).

In the models, we suppose that *Child* (Figure 5 (b)) is an intermediate *worker* that its immediate *supervisor* is one of the LTL operators *and*, *not*, or *until*. Practically, such an actor receives the results of verification from its children, but here, rebec *Child* itself randomly generates this results at line 12, Figure 5 (c). Here, the *Child* randomly generates three results, which is the same three types for a *result message* described above: *true* (modeled with 1), *false* (modeled with 0), and a waiting state (modeled with 2). When modeling a waiting state, message *operator.from_leftChild (false, false)* is used (like line 15 of Figure 5 (c)). The second argument in this message models a waiting state; when this argument is *false*, it shows a waiting state. In this case, the value of the first argument does not matter. But, if the second argument is *true*, it shows that the *Child* is sending a result to its supervisor (*true* or *false*), which is sent via the first argument in the message (see lines 16-27 of Figure 5 (c)).

There are two rebecs of *Child* in our model: *leftWorker* and *rightWorker*. For a binary operator, the worker corresponding to that operator receives two results: one is sent by *rightWorker* and the other is sent by *leftWorker*. As for the unary operator *not*, only messages from *leftWorker* are processed.

In the models, rebec *Supervisor* (Figure 5 (b)) models the immediate *supervisor* of the *worker* whose behavior is supposed to be modeled (i.e. one of the *workers* that acts as one of the LTL operators *not*, *and*, *until*). *Supervisor* using message server *result_fromOp* receives the result of verification from such a *worker* (line 5, Figure 5 (b)). As you can see, this method also has two arguments whose second one models a waiting state.

In Rebeca, verification is performed based on state variables of rebecs so for the rebecs in Figure 5., two variables used to specify properties of our models are state variables *result* and *resultReceived* in rebec *Supervisor* (lines 4-5, Figure 5 (b)). The received result from the LTL operator (*worker*) is saved in variable *result*. We need variable *resultReceived* while modeling because Rebeca initializes state variables at the beginning of execution so the variable *result* has a value even before receiving the real result from the *worker*. Therefore, when *resultReceived* turns into *true*, it denotes that the *Supervisor* has just received the *result* from the *worker* (at line 8 Figure 5 (b)). If the supervisor runs into a waiting state, the *resultReceived* will remain *false*.

### 4.1.1. Modeling the LTL operator *Until*

Figure 6 shows the rebec for the actor that behaves corresponding to LTL operator *until*. As our method uses the Actor model, it is nondeterministic that which actor first processes its incoming messages. Therefore, in the model, you may see some code or state variables for required synchronization.

```
1. reactiveclass SMC (100) {
2.    knownrebecs {
3.       Child leftWorker; Child rightWorker; Until operator;}
4.    statevars { boolean end_of_prog; }
5.    msgsrv initial ( ) { end_of_prog = false; self.sendAP ( ); }
6.    msgsrv sendAP ( ) {
7.   leftWorker.send_result ( );
8.   rightWorker.send_result ( );
9.   end_of_prog = ?(true, false);
10.        if (end_of_prog) operator.endprog ( );
11.   }
12. }
```
**(a)**

```
1. reactiveclass Supervisor (100) {
2.    knownrebecs  {   }
3.    statevars { boolean result;  boolean resultReceived;  }
4.    msgsrv initial ( ) { result = false; resultReceived = false; }
5.    msgsrv result_fromOp (boolean res, boolean st) {
6.   if (st) {
7.       result = res;
8.       resultReceived = true;
9.   }
10. self.reset ( );
11.   }
12.   msgsrv reset ( ) { resultReceived = false; }
13. }
```
**(b)**

```
1. reactiveclass Child (100) {
2.    knownrebecs { Until operator; }
3.    statevars {
4.       byte result;    boolean isLeft;    boolean isRight;
5.    }
6.    msgsrv initial (boolean position) {
7.       result = 2;
8.       if (position) { isRight = true;    isLeft = false; }
9.       else {  isRight = false;    isLeft = true; }
10.   }
11.   msgsrv send_result ( ) {
12.      result = ?(0,1,2);
13.      if (isLeft) {
14.         if (result == 2)
15.            operator.from_leftChild(false, false);
16.         else if (result == 1)
17.            operator.from_leftChild(true, true);
18.         else
19.            operator.from_leftChild(false, true);
20.      }
21.      else {
22.         if (result == 2)
23.            operator.from_rightChild (false, false);
24.         else if (result == 1)
25.            operator.from_rightChild (true, true)
26.      else  operator.from_rightChild(false, true);
27.      }
28.   }
29. }
```
**(c)**

**Figure 5** Rebeca model for the stateless model checker, children and supervisors. (a) The rebec for Stateless *model checker*. (b) The rebec for *Supervisor*. (c) The rebec for *Child*.

For example, when *leftWorker* and *rightWorker* send their own results to rebec *Until*, it is unpredictable that which actor first sends its message. However, we know that both of them send messages about the same state. Therefore, rebec *Until* first requires to receive both of these messages, and then evaluates them. This situation is modeled using state variables *rFlag* and *lFlag* as well as message servers *from_leftChild* and *from_rightChild*.

When both of variables *lFlag* and *rFlag* become *true*, it means that rebec *Until* has received the result from both of its children then it comes to processing. Therefore, method *until_bhv* at line 13 of Figure 6 is executed. In this method, the rebec uses variables *leftOp* and *rightOp*. Variable *leftOp* contains the value of the last message sent by *leftWorker*. In the same way, *rightOp* contains the value of the last message sent by *rightWorker*.

```
1. reactiveclass Until(100) {
2.     knownrebecs { Supervisor parent;  SMC mc; }
3.     statevars {
4.         boolean leftOp;   boolean rightOp;  boolean rFlag;
5.         boolean lFlag;    boolean r_st;     boolean l_st;
6.         boolean updatedRightOp; boolean updatedLeftOp;
7.     }
8.     msgsrv initial ( ) {
9.         leftOp = false;   rightOp = false;   rFlag = false;
10.        lFlag = false;    r_st = false;     l_st = false;
11.        updatedLeftOp = true;   updatedRightOp = false;
12.    }
13.    msgsrv until_bhv ( ) {
14.        if (l_st && r_st) {
15.            if (rightOp)       parent.result_fromOp (true, true);
16.            else if (! leftOp) parent.result_fromOp (false, true);
17.            else if (leftOp && (!rightOp))
18.                parent.result_fromOp (false, false);
19.        }
20.        else if (r_st) {
21.            if (rightOp)       parent.result_fromOp (true, true);
22.            else               parent.result_fromOp (false, false);
23.        }
24.        else if (l_st) {
25.            if (! leftOp)      parent.result_fromOp (false, true);
26.            else               parent.result_fromOp (false, false);
27.        }
28.        else    parent.result_fromOp (false, false);
29.        mc.sendAP( );
30.    }
31.    msgsrv from_leftChild (boolean res, boolean st) {
32.        lFlag = true;    l_st = st;
33.        if (st)    leftOp = res;
34.        if (lFlag && rFlag) {
35.            lFlag = false;     rFlag = false;
36.            updatedRightOp = rightOp;
37.            updatedLeftOp = leftOp;
38.            self.until_bhv( );
39.        }
40.    }
41.    msgsrv from_rightChild (boolean res, boolean st) {
42.        rFlag = true;   r_st = st;
43.        if (st)    rightOp = res;
44.        if (lFlag && rFlag) {
45.            lFlag = false;       rFlag = false;
46.            updatedRightOp = rightOp;
47.            updatedLeftOp = leftOp;
48.            self.until_bhv( );
49.        }
50.    }
51.    msgsrv endprog ( ) {
52.        parent.result_fromOp (true, true);
53.        self.endV ( );
54.    }
55.    msgsrv endV ( ) {   self.endV ();   }
56. }
```

**Figure 6**   Rebeca model for *Until worker*

In order to verify the model, we use two state variables *updatedLeftOp* and *updatedRightOp*, which contain the results respectively sent by *leftWorker* and *rightWorker* after synchronization. This is because the initial state that should be considered to verify the model is not the same initial state in the model. In light of the fact that Rebeca itself initializes state variables, *leftOp* and *rightOp* contains initial values before receiving any messages from *leftWorker* and *rightWorker*. This situation brings about some problem while specifying properties of the model because the *until* operator is sensitive to the initial state of its operands. To resolve this issue, we use auxiliary variables *updatedLeftOp* and *updatedRightOp* for specifying properties.

As mentioned, it is possible for the *until worker* to receive nothing from one or two of its children in practice (i.e. a waiting state). Therefore, it is also possible for the *until* rebec to receive a message that shows a waiting state. Boolean variables *l_st* and *r_st* model this situation. Variable *l_st* holds the second argument of method *from_leftChild*. As mentioned above, if this argument is false, it shows a waiting state. In the same way, variable *r_st* keeps the second argument of method *from_rightChild*. This fact is also true of other *workers* (*Not* and *And*).

Method *until_bhv* models the *until* operator in the $RV^\infty$–$TL_G$ semantic, where an operator evaluates $T_G$ when there is no next state. For this reason, when an iteration of stateless model checking comes to an end, all *workers* will receive the *endprog ( )* message (like line 51 of Figure 6). As this message shows that there will be no next state in the program execution, *workers* send *true* to its *supervisor*. As for *until*, the rebec sends message *result_fromOp (true, true)* to the *supervisor* on line 52.

As the RMC model checker expects non-terminating models so as not to report a deadlock, we use method *endV ( )* in the models.

### 4.1.2. Modeling the LTL operator *And*

The rebec corresponding to LTL operator ∧ in $RV^\infty$–$TL_G$ is shown in Figure 7. This rebec also first receives the results from both of its children using message servers *from_leftChild* and *from_rightChild*, and then it behaves as LTL operator ∧ using method *and_bhv* at line 11 of Figure 7.

```
1. reactiveclass And (100) {
2.   knownrebecs { Supervisor parent; SMC mc; }
3.   statevars {
4.     boolean leftOp;   boolean rightOp;   boolean rFlag;
5.     boolean lFlag;    boolean r_st;      boolean l_st;
6.   }
7.   msgsrv initial ( ) {
8.     leftOp = false;    rightOp = false;   rFlag = false;
9.     lFlag = false;     r_st = false;      l_st = false;
10.  }
11.  msgsrv and_bhv ( ) {
12.     parent.result_fromOp (leftOp & rightOp, true);
13.  }
14.  msgsrv from_leftChild (boolean res, boolean st) {
15.     leftOp = res;   lFlag = true;   l_st = st;
16.     if (rFlag && l_st && r_st) {
17.        lFlag = false;   rFlag = false;   self.and_bhv( );
18.     }
19.     if (rFlag && ((! st) || (! r_st))) {
20.        lFlag = false;      rFlag = false;
21.        parent.result_fromOp (false, false);
22.        mc.sendAP();
23.     }
24.  }
25.  msgsrv from_rightChild (boolean res, boolean st) {
26.     rightOp = res;   rFlag = true;   r_st = st;
27.     if (lFlag && l_st && r_st) {
28.        lFlag = false;      rFlag = false;
29.        self.and_bhv( );
30.     }
31.     if (lFlag && ((! st) || (! l_st))) {
32.        lFlag = false;      rFlag = false;
33.        parent.result_fromOp (false, false);
34.        mc.sendAP();
35.     }
36.  }
37.  msgsrv endprog( ){
38.     if ((!r_st) && (!l_st))
39.        parent.result_fromOp (true, true);
40.     else if (! l_st)
41.        parent.result_fromOp (rightOp, true);
42.     else if (! r_st)
43.        parent.result_fromOp (leftOp, true);
44.     self.endV ( );
45.  }
46.  msgsrv endV ( ) {  self.endV ( );   }
47. }
```

**Figure 7**   Rebeca model for *And worker*

Rebec *And* uses two state variables *leftOp* and *rightOp* for saving the results sent by *leftWorker* and *rightWorker*, respectively. In addition, these variables are used for specifying the properties of the model as well as its verification.

### 4.1.3. Modeling the LTL operator *Not*

The rebec shown in Figure 8 models the behavior of the *worker* that acts as LTL operator ¬ in $RV^\infty$–$TL_G$. For the sake of brevity, we use the same structure of *Child* and *SMC* shown in Figure 5 for the *Not* rebec. Therefore, this rebec also has a message server named *from_righChild*, while this message server has no effect on the behavior of this rebec because it only considers the result sent by *leftWorker* saving it in state variable *opr* at line 10. This state variable is used for verifying the model as well.

After receiving the message from its child, rebec *Not* processes that using method *not_bhv* at line 5. That is, it negates *opr* and sends it for its *Supervisor* (i.e. variable *parent*) on line 6.

```
1. reactiveclass Not (100) {
2.    knownrebecs {  Supervisor parent;  SMC mc;  }
3.    statevars {   boolean opr;    }
4.    msgsrv initial ( ) {    opr = false;    }
5.    msgsrv not_bhv (boolean opr) {
6.       parent.result_fromOp (!opr, true);
7.    }
8.    msgsrv from_leftChild (boolean res, boolean st) {
9.       if (st) {
10.         opr = res;
11.         self.not_bhv(opr);
12.      }
13.      else {
14.         parent.result_fromOp (false, false);
15.         mc.sendAP();
16.      }
17.   }
18.   msgsrv from_rightChild (boolean resm, boolean st) {
         // There is no right child!
         // this message will never be used.
19.   }
20.   msgsrv endprog ( ) {
21.      parent.result_fromOp (false, true);
22.      self.endV ( );
23.   }
24.   msgsrv endV ( ) {   self.endV ();   }
25. }
```

**Figure 8**   Rebeca model for *Not worker*

## 4.2. Verification results

To verify our model, we use model checking; in this section, properties that should be satisfied by models are specified in LTL. We have used model checker RMC [32] for verifying our models.[§] In terms of LTL operator *U*, the safety property, which was verified and proved to be true in the model, is that when left operand remains *true* until the right operand becomes *true*, *Supervisor* should receive a *true* message from rebec *Until*, otherwise it should receive a *false*. In our method, as described above, the left operand is the same result sent by *leftWorker* saved in *updatedLeftOp*, and the right operand is the same result sent by *rightWorker* saved in *updatedRightOp*. This property is specified in LTL as follows:

$G$ ( (*until.updatedLeftOp* U *until.updatedRightOp*) ∧ *parent.resultReceived*) → *parent.result*)

$G$ (¬ (*until.updatedLeftOp* U *until.updatedRightOp*) ∧ *parent.resultReceived*) → ¬ *parent.result*)

The safety property that rebec *And* is expected to hold is that *Supervisor* receives a *true* from rebec *And* when both of its left operand and right operand are *true*, otherwise it should receive a *false*. This property was also verified and proved true. In our model, the left operand is the same result sent by *leftWorker*, and the right operand is one sent by *rightWorker*, which are saved in variables *leftOp* and *rightOp*, respectively. Therefore, the LTL specification of this property is as follows:

$G$ ( (*and.leftOp* ∧ *and.rightOp*) ∧ *parent.resultReceived*) → *parent.result*)

$G$ ( ¬ (*and.leftOp* ∧ *and.rightOp*) ∧ *parent.resultReceived*) → ¬ *parent.result*)

---

[§] Rebeca models as well as the output of have been attached to this paper.

For LTL operator ¬, it is expected that *Supervisor* receives a true from rebec *Not* if the operand of operator ¬ is *false*. Obviously, in our method, the operand of a *not* operator is an actor. In the models, the value of this operand is the same result sent by *leftWorker* to rebec *Not*. The rebec saves this value into its state variable *opr*. Therefore, the LTL property should be held by rebec *Not* is as follows:

$$G ((\neg \, not.op \wedge \, parent.resultReceived) \rightarrow parent.result)$$
$$G ((not.op \wedge \, parent.resultReceived) \rightarrow \neg \, parent.result)$$

Table 2 shows the results of the verification of the forgoing properties by model checker RMC [32].

**Table 2.** The verification results

| Property | RMC Version 2.2 | | | | | |
|---|---|---|---|---|---|---|
| | **Status** | **Depth reached** | **Transitions** | **States** | **Time (sec)** | **Memory (MB)** |
| **Until** | bound reached | 10,000 | 14,621 | 10,019 | 0 | 722.062 |
| **And** | bound reached | 10,000 | 10,262 | 9,912 | 1 | 3,015.433 |
| **Not** | bound reached | 10,000 | 10,262 | 9,912 | 1 | 3,014.229 |

## 5. An illustrative example

This section describes a simple example of stateless model checking of an LTL property to illustrate the proposed method. This example is a version of the mutual exclusion problem with two threads. The pseudo code of the problem is shown in Figure 9 (a). The safety property that program should satisfy is that two threads do not enter the critical section at the same time, which is specified in LTL as "$G (\neg (crit1 \wedge crit2))$"; consequently, the *APs* used by the user in the LTL property are *crit1* and *crit2*. In this section, we describe the process of verification of this property step by step. It should be pointed out that this property can be verified via the $RV^\infty$– $TL_G$ semantic because its grammar corresponds to the $TL_G$ category.

As we described, the stateless model checker is expected to partition the program code according to visible operations. You can see the partitioned code of Figure 9 (a) in Figure 9 (b). Based on the rule of partitioning, each of *T1* and *T2* is divided into six locations. During stateless model checking, the model checker schedules threads based on these locations.

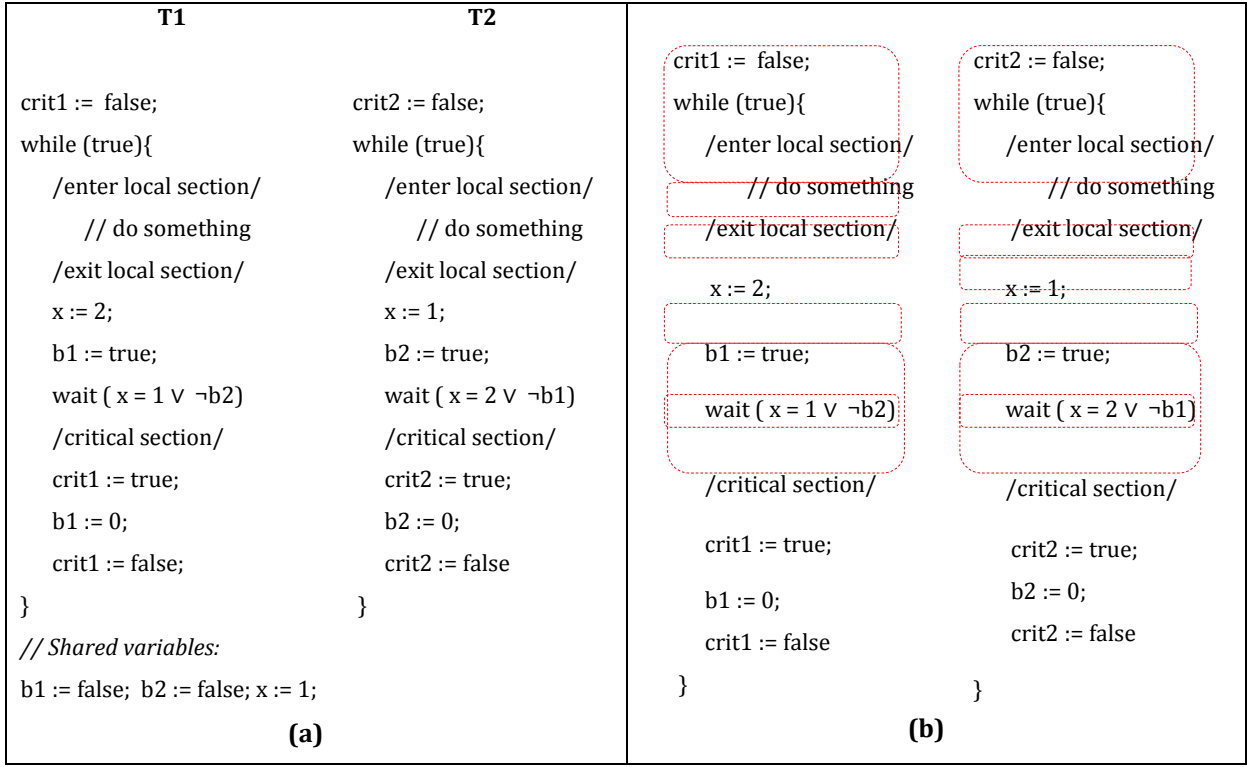

**Figure 9**  An example of the mutual exclusion problem

At the end of each location (i.e. after a thread yields the CPU due to reaching the end of its current location), the model checker informs *condition checkers* about the current status of their *AP*. Hereafter, we suppose that stateless model checker has partitioned the program (i.e. Figure 9 (b)) and is ready to start model checking.

Now, let us start explaining the verification process for this example. First of all, the *master* actor loads the defined property. Here, the user has defined only one property so *master* only creates one *property checker*, which is responsible for verifying the specified property. Now, *property checker* should create the hierarchy of *workers*. Frist, it standardizes the specified property as follows:

$$G\,(\neg\,(crit1 \wedge crit2)) = \neg F\,\neg\,(\neg\,(crit1 \wedge crit2)) = \neg\,(true\ U\ (crit1 \wedge crit2))$$

Next, *property checker* should parse this property and initiates creation of the hierarchy of *workers*. The parse tree for this property is shown in Figure 10. After parsing, *property checker* creates a *worker* that behaves as a *not* operator, and sends the sub-tree of the *not* operator to this *worker*. This *worker* also parses the received sub-tree, creates an *until worker*, and sends the sub-tree under operator *until* to the created *worker*. The *until worker* also parses its sub-trees and creates a *condition checker* as its right child, which only generates *true*. For its left child, the *until worker* creates an *and worker* sending the reminder of the tree to this *worker*. After parsing the received sub-tree, the *and worker* creates two *condition checkers* that check *APs crit1* and *crit2*.

After creating the hierarchy, the model checker starts to explore the state space and verification. Suppose the model checker first schedules *T2*; therefore, *T2* performs its computation from the beginning of location 1 to the end of this location. At the end of this location, *T2* is preempted, and the stateless model checker sends the status of *APs* to the *condition checkers*. At this time, both of *crit1* and *crit2* are *false*, hence "*condition checkers 2*" and "*condition checkers 3*" send false to the "*worker 3*" (see Figure 10). As "*worker 3*" is an *and* operator, it generates *false* because both of its operands are *false*. Therefore, "*worker 2*" receives a *false* message from its right *worker* and receives a *true* from its left *worker*.

According to the behavior of an *until worker* (Figure 6), the "*worker 2*" still waits for hearing from its children in the next status. For the sake of brevity, we summarize this process in Table 3. As you can see in Table 3, the model checker schedules threads as follows: "$s_1$: T2, $s_2$: T2, $s_3$: T1, $s_4$: T1, $s_5$: T1, $s_6$: T1, $s_7$: **T1**, $s_8$: T2, $s_9$: T2, $s_{10}$: T2". The described situation recurs until $s_7$.

We go on explaining with $s_7$, where *T1* enters the critical section and *crit1* becomes *true*. As a result, at the end of this state, "*condition checker 2*" receives a *true* message and "*condition checker 3*" receives a *false*. Consequently, "*worker 3*" generates a *false* message so the reminder of the process is similar to what was described above. This situation recurs until the end of $s_9$, where *T2* also enters the critical section causing *crit2* to become *true*.

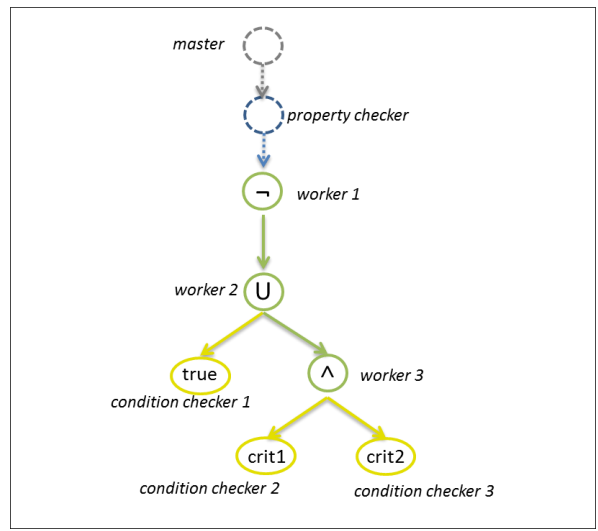

**Figure 10**   The parse tree of property ¬ (*true* U (*crit1* ∧ *crit2*))

At the end of $s_{10}$, the model checker sends *true* to both "*condition checker 2*" and "*condition checker 3*", whereby "*worker 3*" also concludes a *true* result, and sends it to "*worker 2*". Therefore, "*worker 2*" receives *true* from both left *worker* and right *worker* so it also sends a *true* message to "*worker 1*". As "*worker 1*" is a *not* operator, it negates the received result. Consequently, "*worker 1*" evaluates the result of verification as *false*. This result is sent to *property checker*. When *property checker* receives a (*false, true*) message, it concludes that a violation has occurred.

To briefly explain this example without complexity, we considered a non-terminating program. But, suppose that loop "*while (true)*" does not exist. Therefore, the program eventually comes to an end. In this case, the stateless model checker generates possible finite executions of the program. Suppose it generates different executions as follows:

$i_{0,s_0}$ : *T1, T1, T2, T1, T1, T2, T2, T2, T1, T1, T2, T2, T2.*

$i_{1,s_0}$ : *T1, T2, T2, T1, T2, T1, T1, T2, T2, T2, T1, T1, T1.*

...

$i_{k,s_0}$ : *T2, T2, T1, T1, T1, T1, T1, T2, T2, T2.*

As a result, the property is independently checked in each iteration. If the property is satisfied in all iterations, then the program satisfies the property. Obviously, if the property is violated in (at least) one iteration, for example in $i_{k,s_0}$, it means that the program does not satisfy the property.

## 6. Implementation issues

All actors of the verification hierarchy act in parallel with the stateless model checker. This hierarchy is very quickly formed at the beginning of stateless model checking. We are going to implement our method by Erlang programming language [31], in which processes (i.e. actors) are very lightweight and cheap to create (about 100 times lighter than threads) [53]. Message passing in Erlang is also very fast (about one micro second) [30, 53]. Therefore, there is no concern about the process creation and message-passing overhead. Erlang provides the best implementation of the Actor model [28], whereby we can precisely implement the proposed method.

**Table 3.** The verification process for the example shown in Figure 9

| | $s_0$ | $s_1$ | $s_2$ | $s_3$ | $s_4$ | $s_5$ | $s_6$ | $s_7$ | $s_8$ | $s_9$ | $s_{10}$ |
|---|---|---|---|---|---|---|---|---|---|---|---|
| **The scheduled thread** | - | T2 | T2 | T1 | T1 | T1 | T1 | T1 | T2 | T2 | T2 |
| **Location number *T1 points to*** | 1 | 1 | 1 | 2 | 3 | 4 | 5 | 6 | 6 | 6 | 6 |
| **Location number *T2 points to*** | 1 | 2 | 3 | 3 | 3 | 3 | 3 | 3 | 4 | 5 | 6 |
| **x1** | 1 | 1 | 1 | 2 | 2 | 2 | 2 | 2 | 2 | 2 | 2 |
| **b1** | false | false | false | false | true | true | true | true | true | true | true |
| **b2** | false | false | false | false | false | false | false | false | true | true | true |
| **crit1** | false | false | false | false | false | false | false | true | true | true | <span style="color:red">true</span> |
| **crit2** | false | false | false | false | false | false | false | false | false | false | <span style="color:red">true</span> |
| **Message received by "*condition checker 2*"** | - | false, true | false, true | false, true | false, true | false, true | false, true | true, true | true, true | true, true | true, true |
| **Message received by "*condition checker 3*"** | - | false, true | false, true | false, true | false, true | false, true | false, true | false, true | false, true | false, true | true, true |
| **Message sent by "*worker 3*"** | - | false, true | false, true | false, true | false, true | false, true | false, true | false, true | false, true | false, true | true, true |
| **Message sent by "*worker 2*"** | - | false, false | false, false | false, false | false, false | false, false | false, false | false, false | false, false | false, false | true, true |
| **Message sent by "*worker 1*"** | - | false, false | false, false | false, false | false, false | false, false | false, false | false, false | false, false | false, false | **<span style="color:red">false, true</span>** |

## 7. Conclusions

This paper proposes a new verification method for stateless model checking of LTL properties. In our method verifies formulae dynamically without storing any program states. The proposed method is designed based on the Actor model. Thanks to this model, we can create cheap and lightweight actors that check LTL properties simultaneously with stateless state space exploration.

The method proposed in this paper is designed as the unit of LTL checking for DSCMC [27, 35], which is a parallel stateless code model checker. We are implementing this method in DSCMC.

This tool needs to analyze and instrument program code before performing stateless model checking. Currently, code is manually instrumented in DSCMC. Therefore, in the future, code instrumentation must be automated for using DSCMC in large programs. Once this has been done, we will be able to utilize the proposed method for real-world programs.

As for stateless model checking, it may be impractical to precisely handle non-deterministic user input. Then in practice, using the method to verify large programs may be transformed to *systematically* testing, but it is still powerful enough to explore the state space of large programs whose state space exploration is impractical using state-based methods [1, 4, 25, 26, 37, 54]. However, to cover more execution paths, the method can be improved by employing test generation techniques, such as white-box fuzz testing [55, 56] and symbolic execution [57].

## Acknowledgement

We are grateful to Iran National Science Foundation (INSF) for financial support of this research.